\documentclass[a4paper,11pt]{article}
\usepackage{pos}
\usepackage{slashed}

\title{Non-diagonal DVCS with spinless hadron-to-two-hadron transition $\gamma^*\pi\to\gamma\pi\pi$}

\author*[a]{Sangyeong Son}

\affiliation[a]{Department of Physics, Kyungpook National University,\\
 41566, Daegu, Republic of Korea}

\emailAdd{thstkd3754@gmail.com}

\abstract{
We present the formalism of hadron-to-two-hadron transition generalized parton distributions (GPDs) for spinless hadron case.
Definitions of the twist-2 unpolarized and polarized $\pi\to\pi\pi$ transition GPDs are introduced with a particular choice of kinematic variables that characterize the produced two-pion system.
In the vicinity of $\rho(770)$, we work out the two-pion decay angular distributions of the $e^-\pi\to e^-\gamma\pi\pi$ cross section, incorporating both the Bethe-Heitler and deeply virtual Compton scattering processes. Each cross section exhibits a distinctive angular distribution, which is sensitive to the polarization states of the produced $\rho(770)$ resonance. In addition, we construct the double partial-wave expansion of $\pi\to\pi\pi$ transition GPDs in the two-pion decay angles as a generalization of GPDs for transition from a pion to a spin-$\ell$ resonance state. With the help of the Omnès representation, we constrain the $\pi\to\pi\pi$ transition GPDs in terms of the low energy $\pi\pi$ scattering phase shift and build a phenomenological model for these GPDs.
}

\FullConference{The 21st International Conference on Hadron Spectroscopy and Structure (HADRON2025)\\
 27 - 31 March, 2025\\
Osaka University, Japan\\}

\begin{document}
\maketitle

\section{Introduction}
Non-diagonal hard exclusive reactions, which involve transitions from a target hadron to excited states, extend the exploration of hadron structure beyond ground-state nucleons to include baryon resonances.
Under QCD factorization, these processes allow for the introduction of generalized parton distributions (GPDs) for transitions such as $N\to N^*$.
As a natural extension of the GPD framework, transition GPDs provide a means to probe the internal QCD structure underlying hadronic excitations, offering insight into the dynamical features of resonance formation in terms of quarks and gluons.
A recent white paper~\cite{Diehl:2024bmd} provides a review of the current status of theoretical achievements and experimental prospects for studying hadron resonance structure through non-diagonal hard exclusive reactions within the framework of transition GPDs.

A unified description of baryon resonances can be developed through the formulation of $N \to \pi N$ transition GPDs~\cite{Polyakov:1998sz, Polyakov:2006dd}. These objects generalize the concept of nucleon-to-resonance transition GPDs across the complete $\pi N$ resonance region and allow for their systematic representations via analytic continuation in the $\pi N$ invariant mass. Given the two-body nature of the final state, however, the $N \to \pi N$ GPDs possess more richer kinematical structure than in the diagonal case. Hence, it is useful to begin with a simplified example involving only spinless hadrons--namely, the $\pi\to\pi\pi$ transition GPDs.

This proceedings contribution summarizes the results of our recent study on the $\pi\to\pi\pi$ transition GPD framework, originally presented in Ref.~\cite{Son:2024uxa}.

\section{Non-diagonal hard exclusive reaction with $\pi\to\pi\pi$ transition GPDs}
\begin{figure}[b]
    \centering
    \includegraphics[width=0.6\linewidth]{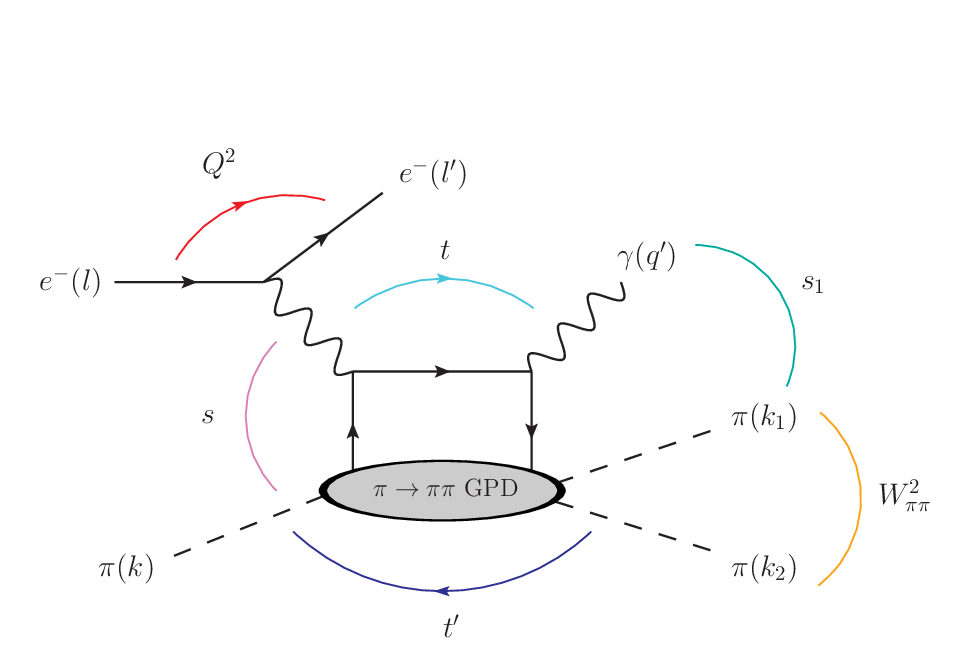}
    \caption{Factorization mechanism of the DVCS process with $\pi\to\pi\pi$ transition GPDs. The invariant variables relevant to the kinematics of the $e^-\pi\to e^-\gamma\pi\pi$ process are indicated. The diagram with crossed virtual and real photon lines is not shown explicitly.}
    \label{fig:ND_DVCS}
\end{figure}
As a spinless hadron example of non-diagonal hard exclusive reaction, we consider the deeply virtual Compton scattering (DVCS) $\gamma^*\pi\to\gamma\pi\pi$. This process serves as a subprocess of photon electroproduction involving the $\pi\to\pi\pi$ transition, which is depicted in Fig.~\ref{fig:ND_DVCS}:
\begin{eqnarray}
    e^-(l)+\pi(k) \to e^-(l')+ \gamma(q') + \pi(k_1) + \pi(k_2).
\end{eqnarray}
The collinear factorization theorems~\cite{Collins:1996fb, Collins:1998be} are assumed to be valid for non-diagonal hard exclusive process involving transitions from a hadron to a two-hadron system with low invariant mass. This enables us to introduce the $\pi\to\pi\pi$ transition GPDs in the generalized Bjorken kinematics, where $s = (k+q)^2$, $s_1 = (q'+k_1)^2$, and $Q^2 = -q^2 = -(l-l')^2$ are large; the Bjorken variable $x_B = \frac{Q^2}{2k\cdot q}$ and $\frac{s_1}{s}$ are fixed; and $t = (q-q')^2$, $t' = (k_2-k)^2$, and $W_{\pi\pi}^2$ are of hadronic mass scale.
The twist-2 unpolarized and polarized $\pi\to\pi\pi$ transition GPDs are defined through the Fourier transform of the non-diagonal hadronic matrix elements of the non-local light cone quark operators,
\begin{eqnarray}
    && \int \frac{d\lambda}{2\pi}e^{i\lambda x (\bar{P}\cdot n)} \sum_{q}e_q^2\big \langle \pi(k_1)\pi(k_2) \big| \bar{q}\left(-\frac{\lambda n}{2}\right)\gamma\cdot n
    \begin{Bmatrix}
        1 \\ \gamma_5\\
    \end{Bmatrix}
    q\left( \frac{\lambda n}{2}\right) \big | \pi(k) \big \rangle \nonumber \\
    &&= \frac{1}{\bar{P}\cdot n}
    \begin{Bmatrix}
        \frac{i\varepsilon(n,\Bar{P},\Delta,k_1)}{f_\pi^3} H^{\pi\to\pi\pi} \\
        \frac{i}{f_\pi}\tilde{H}^{\pi\to\pi\pi}
    \end{Bmatrix}
    (x,\xi,t;W_{\pi\pi}^2,\theta_\pi^*,\varphi_\pi^*),
\end{eqnarray}
where $\bar{P} = \frac{k+k_1+k_2}{2}$ is the average hadron momentum, $n \propto (1,0,0,-1)$ is the light cone vector, $\varepsilon_{\mu\nu\rho\sigma}$ is the totally anti-symmetric tensor\footnote{We used the shorthand notation $\varepsilon(a,b,c,d)=\varepsilon_{\mu\nu\rho\sigma}a^\mu b^\nu c^\rho d^\sigma$ with the convention $\varepsilon_{0123} = -\varepsilon^{0123}=+1$.}
, and $f_\pi \simeq 92.4~\mathrm{MeV}$ is the pion decay constant. 

It is worth noting that the GPDs for hadron-to-two-hadron transition possess more arguments than the usual GPDs. The $\pi\to\pi\pi$ GPDs depend not only on the usual GPD variables--longitudinal parton momentum fraction $x$, the skewness variable $\xi = -\frac{\Delta\cdot n}{2\bar{P}\cdot n}$, and the momentum transfer $t$--but also on three additional variables that characterize the final state two-pion system. One of these is the $2\pi$ invariant mass $W_{\pi\pi}^2$. In order to utilize these transition GPDs in a form suitable for PW analysis and application of dispersive techniques, we choose the polar and azimuthal angles of the $2\pi$ decay, $\Omega_\pi^* = (\theta_\pi^*, \varphi_\pi^*)$, defined in the helicity frame ($\vec{k}_1= -\vec{k}_2$)~\cite{Byckling1973}, as the remaining two variables.

It is instructive to investigate the $2\pi$ decay angular distribution of the cross sections associated with the Bethe-Heitler (BH) and the DVCS mechanisms. 
The sevenfold differential cross section of the $e^-\pi\to e^-\gamma\pi\pi$ reaction reads
\begin{eqnarray}
        && \frac{d^7\sigma}{dx_B dQ^2 dt d\Phi dW_{\pi\pi}^2 d\Omega_\pi^*} = \frac{1}{256(2\pi)^7}\frac{x_B y^2}{Q^4 \sqrt{1+\frac{4m_\pi^2 x_B^2}{Q^2}}}\sqrt{1-\frac{4m_\pi^2}{W_{\pi\pi}^2}} \bar{\sum_i}\sum_f \left|\mathcal{M}_{\mathrm{BH}}+\mathcal{M}_{\mathrm{DVCS}}\right|^2, \nonumber \\
    \label{CS_7fold}
\end{eqnarray}
where $\Phi$ denotes the angle between the leptonic plane and the production plane, and we account for the isolated $\rho(770)$ resonance contribution to the corresponding amplitude,
\begin{eqnarray}
    \mathcal{M}(e^-\pi\to e^-\gamma\rho \to e^-\gamma\pi\pi) &=& C_\mathrm{iso}g_{\rho\pi\pi}(k_1-k_2)_\mu \frac{i\sum_{\lambda_\rho} \mathcal{E}^\mu(p_{\pi\pi}, \lambda_\rho) \mathcal{E}^{*\nu}(p_{\pi\pi}, \lambda_\rho) }{W_{\pi\pi}^2 - m_\rho^2 + im_\rho\Gamma_\rho} \nonumber \\
    && \mbox{} \times \mathcal{M}_\nu(e^-\pi\to e^-\gamma\rho).
    \label{Amp_pi-2pi}
\end{eqnarray}
Here, $C_{\mathrm{iso}} = 1/\sqrt{2}$ for $\rho^+\to\pi^+\pi^0$, $m_\rho = 770~\mathrm{MeV}$ is the $\rho$-meson mass, $\Gamma_\rho \simeq 149.1~\mathrm{MeV}$~\cite{ParticleDataGroup:2022pth} is the $\rho\to\pi\pi$ decay width, $g_{\rho\pi\pi} \simeq 6.01$ is the $\rho\pi\pi$ effective coupling, and $\mathcal{E}(p_{\pi\pi},\lambda_\rho)$ represents the polarization vector of $\rho$ with momentum $p_{\pi\pi}$ and $\lambda_\rho$ polarization state. Integrating the amplitude squared over the azimuthal angle of $2\pi$ decay results in the $\theta_\pi^*$ distribution of the cross section,
\begin{eqnarray}
    &&\int_0^{2\pi}d\varphi_\pi^* \big|\mathcal{M}(e^-\pi\to e^-\gamma\rho\to e^-\gamma\pi\pi)\big|^2 
    = C_\mathrm{iso}^2 g_{\rho\pi\pi}^2\frac{W_{\pi\pi}^2-4m_\pi^2}{(W_{\pi\pi}^2 - m_\rho^2)^2 + m_\rho^2\Gamma_\rho^2}\frac{4\pi}{3} \nonumber \\ 
    && \quad\times \sum_{\lambda_\rho} \big|\mathcal{M}\left(e^-\pi\to e^-\gamma\rho(p_{\pi\pi}, \lambda_\rho)\right)\big|^2 \biggl[\frac{3}{2}\cos^2\theta_\pi^* \delta_{\lambda_\rho, 0} + \frac{3}{4}\sin^2\theta_\pi^*\left(\delta_{\lambda_\rho, 1} + \delta_{\lambda_\rho, -1}\right)\biggr],
    \label{Polar_angle_distribution}
\end{eqnarray}
which exhibits distinctive angular structures depending on the polarization state of the intermediate resonance. 
\begin{figure}[t]
    \centering
    \includegraphics[width=0.45\linewidth]{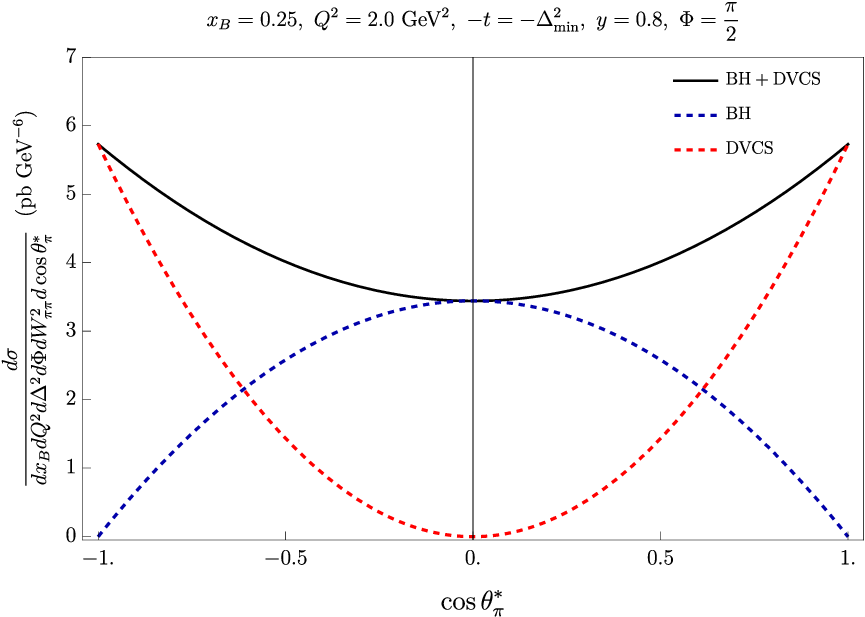}\qquad
    \includegraphics[width=0.45\linewidth]{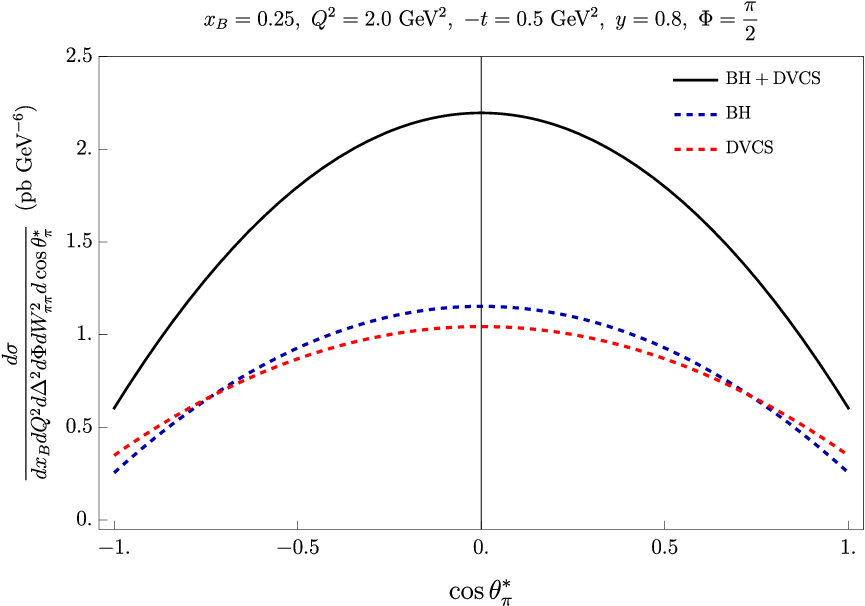}
    \caption{The $\cos\theta_\pi^*$ distributions of the $e^-\pi^+\to e^-\gamma\rho^+ \to e^-\gamma \pi^+\pi^0$ differential cross section. The figures are taken from Ref.~\cite{Son:2024uxa}.}
    \label{fig:cross_BH-DVCS}
\end{figure}
In Fig.~\ref{fig:cross_BH-DVCS}, we show the $2\pi$ decay polar angular distribution of the differential cross section of $e^-\pi^+\to e^-\gamma \pi^+ \pi^0$, incorporating both the BH and DVCS mechanisms in the vicinity of $\rho(770)$. The estimations are performed for $x_B = 0.25$, $Q^2 = 2.0~\mathrm{GeV}^2$, $y = 0.8$, and $\Phi = \frac{\pi}{2}$ which are typical kinematic conditions of the JLab@12GeV facility. At the minimum value of $-t_{\mathrm{min}} \simeq 0.2~\mathrm{GeV}^2$, the total cross section is dominated by the longitudinally polarized $\rho(770)$, as can be seen from Eq.~(\ref{Polar_angle_distribution}). With increasing $-t$, the DVCS cross section becomes dominated by the circular polarization states of the intermediate $\rho(770)$ resonance, as reflected in the change of its curvature to the negative sign. In contrast, the angular distribution of BH mechanism indicates that the dominance of the $\lambda_\rho = \pm 1$ polarization states remains unchanged with varying $-t$. The clear distinction between the BH and DVCS behaviors in the $2\pi$ decay angle implies the utility of this formalism in distinguishing these mechanisms.


\section{Dispersive analysis on the $\pi\to\pi\pi$ transition GPDs}

\begin{figure}[t]
    \centering
    \includegraphics[width=0.5\linewidth]{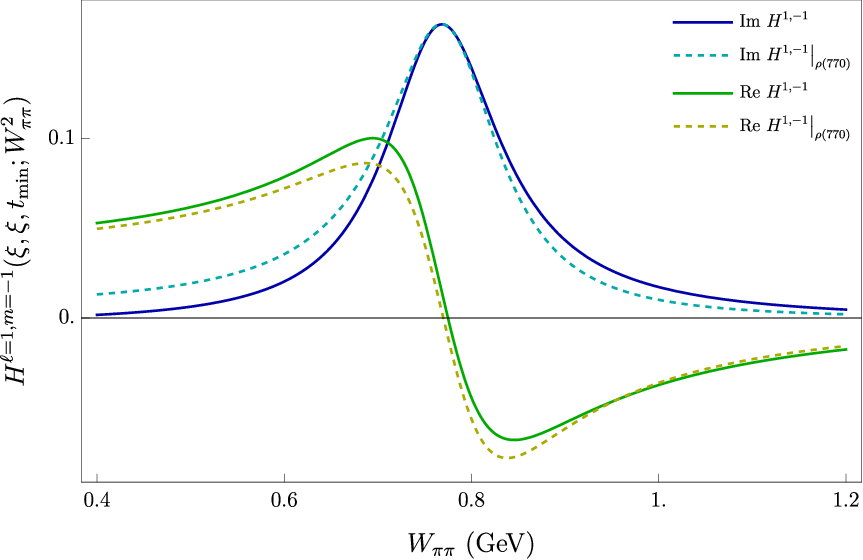}
    \caption{The $W_{\pi\pi}$ distribution of the real and imaginary parts of the PW GPD $H^{\ell = 1, m=-1}(\xi,\xi,t;W_{\pi\pi}^2)$ on the crossover line $x = \xi \simeq 0.15$ and $t = t_{\min}$. This figure is taken from Ref.~\cite{Son:2024uxa}.}
    \label{fig:PW_GPD}
\end{figure}

The generalization of Eq.~(\ref{Amp_pi-2pi}) for a $2\pi$-resonance with spin-$\ell$ leads us to construct the partial-wave (PW) expansion of $\pi\to\pi\pi$ transition GPDs in the real-valued spherical harmonics of the $2\pi$ decay angles,
\begin{eqnarray}
   H^{\pi\to\pi\pi}(x,\xi,t;W_{\pi\pi}^2, \theta_\pi^*, \varphi_\pi^*) &=& \frac{1}{ \sin\theta_\pi^* |\sin\varphi_\pi^*| }\sum_{\ell = 1}^\infty\sum_{m = -\ell}^{-1}H^{\ell,m}_{}(x,\xi,t;W_{\pi\pi}^2)\operatorname{Y}_{\ell, m}(\theta_\pi^*, \varphi_\pi^*),  \\
   \tilde{H}^{\pi\to\pi\pi}(x,\xi,t;W_{\pi\pi}^2, \theta_\pi^*, \varphi_\pi^*) &=& \sum_{\ell = 0}^\infty\sum_{m = 0}^\ell \tilde{H}^{\ell,m}_{}(x,\xi,t;W_{\pi\pi}^2)\operatorname{Y}_{\ell, m}(\theta_\pi^*, \varphi_\pi^*). \label{PW_expansion_GPDs}
\end{eqnarray}
Following the approach formulated in Refs.~\cite{Polyakov:1998ze, Lehmann-Dronke:1999vvq, Lehmann-Dronke:2000hlo}, we apply dispersive methods to the $\pi\to\pi\pi$ GPDs to construct a unified description of pion-to-resonance transition in the complete $2\pi$-resonance region.
We make use of the above PW expansion with the application of the Watson-Migdal final-state interaction theorem~\cite{Watson:1954uc, WOS:A1955WN97000001}
\footnote{We note that, unlike the usual GPDs, $\pi\to\pi\pi$ GPD above the $2\pi$-threshold, $W_{\pi\pi}^2 \geq 4m_\pi^2$, acquires the imaginary part, given by the discontinuity along the cut in the complex $W_{\pi\pi}^2$-plane.}. Its result can then be rewritten by the Muskhelishvili-Omnès type $N$-subtracted dispersion relation~\cite{Omnes:1958hv},
\begin{eqnarray}
&&H^{\ell, m} (x,\xi,t; W^2_{\pi \pi}) \nonumber \\
&&=\sum_{n=0}^{N-1} \frac{W^{2 n}_{\pi \pi}}{n !} \frac{d^n}{dW_{\pi \pi}^{2 n}} H^{\ell, m} (x,\xi,t; W^2_{\pi \pi})\biggr|_{W^2_{\pi\pi}=0}+ \frac{W^{2 N}_{\pi \pi}}{\pi} \int_{4 m_\pi^2}^{\infty}d\omega \frac{\mathrm{Im}H^{\ell,m}(x,\xi,t;\omega)}{\omega^N\left(\omega-W^2_{\pi \pi}-i \epsilon\right)}.
\end{eqnarray}
The solution of the $N = 1$ subtracted dispersion relation is given by the Omnès representation with the PW GPD $H^{\ell,m}$ at the threshold,
\begin{eqnarray}
    H^{\ell, m}(x,\xi,t;W_{\pi\pi}^2) = H^{\ell,m}(x,\xi,t;W_{\pi\pi}^2=0)\exp\left[\frac{W_{\pi\pi}^2}{\pi}\int_{4m_\pi^2}^\infty d\omega \frac{\delta^{I=1}_\ell(\omega)}{\omega(\omega-W_{\pi\pi}^2-i\epsilon)}  \right], \label{Omnes}
\end{eqnarray}
in which the $2\pi$ invariant mass dependence of $H^{\ell,m}$ is constrained to the low energy $\pi\pi$ scattering phase shifts $\delta_l^{I}(\omega)$.
In Fig.~\ref{fig:PW_GPD}, we present the PW GPD $H^{\ell=1,m=-1}$ constructed using the Omnès representation in Eq.~(\ref{Omnes}), incorporating the $P$-wave ($\ell = 1$) $\pi\pi$ scattering phase shift dominated by the isolated $\rho(770)$. The $x$- and $\xi$- dependencies of this GPD are modeled using the Radyushkin double distribution Ansätz~\cite{Radyushkin:1998bz, Radyushkin:1998es}, while the $t$-dependence is introduced through a factorized dipole parametrization. The constructed GPD at the threshold is then fitted to the Breit-Wigner form of the $\rho(770)$ resonance contribution in Eq.~(\ref{Amp_pi-2pi}). Notably, this Omnès-based description can provide a systematic framework for modeling PWs that are not solely governed by isolated Breit-Wigner resonances.

\section{Summary}
In this proceedings contribution, we briefly address the non-diagonal DVCS $\gamma^*\pi\to\gamma\pi\pi$ in the generalized Bjorken kinematics, introducing the concept of $\pi\to\pi\pi$ transition GPDs that provide a unified description of pion-to-resonance GPDs. In the vicinity of $\rho(770)$, we estimate the $2\pi$ decay angular distribution of the BH and DVCS contributions to the $e^-\pi^+\to e^-\gamma\pi^+\pi^0$ cross section at the kinematics of JLab@12GeV. This observable is shown to be sensitive to the polarization states of the intermediate resonance and exhibit distinctive angular structures for each mechanism, depending on the kinematic conditions.
With help of the PW expansion of the $\pi\to\pi\pi$ transition GPDs and dispersive analysis, we construct a phenomenological model for the PW GPD, $H^{\ell = 1, m = -1}$, with constraining its $W_{\pi\pi}^2$ dependence using low energy $\pi\pi$ scattering data.
Further details on the construction of this framework, including the application of the Froissart-Gribov projection technique are presented in Ref.~\cite{Son:2024uxa}.
The generalization for the $N \to \pi N$ transition across the $\pi N$ resonance region can be useful in the studies of baryon resonances through non-diagonal hard exclusive reactions.

\acknowledgments
This work was supported by Basic Science Research Program through the National Research Foundation of Korea (NRF) funded by the Ministry of Education RS-2023-00238703 and RS-2018-NR031074.

\bibliographystyle{JHEP}
\bibliography{main}

\end{document}